\documentclass[conference,10pt]{IEEEtran}
\usepackage{amsmath,graphicx}

\graphicspath{{./figures/}}
\DeclareGraphicsExtensions{eps, pdf, jpg}

\usepackage{url}

\usepackage{amsfonts}
\usepackage{amssymb}
\usepackage{amsbsy}
\usepackage{times}
\usepackage{default}
\usepackage{color}
\usepackage[noadjust]{cite}
\usepackage{enumerate}
\usepackage{tikz}
\usepackage{subcaption}
\usepackage{amsthm}
\usepackage{dsfont}

\usepackage[binary-units=true]{siunitx}

\usetikzlibrary{arrows}

\usepackage{multirow}
\usepackage{booktabs}
\usepackage{textcomp}

\usepackage{hyperref}

\newcommand{\figref}[1]{{Fig.~\ref{#1}}}

\usepackage{footnote}
\makesavenoteenv{tabular}
\makesavenoteenv{table}
\usepackage{pgf,tikz}

\usepackage{pgfplots}
\pgfplotsset{compat=1.13}

\usepackage[font=small]{caption}

\usepackage{float}

\newcommand{\ChipArea}{0.258\,mm$^2$}
\newcommand{\AluArea}{0.07\,mm$^2$}
\newcommand{\maxClkFreq}{413\,MHz}
\newcommand{\avgPower}{210\,{\textmu}W/MHz}
\newcommand{\avgPoweru}{210\,uW/MHz}
\newcommand{\TestVolt}{1.2\,V}
\newcommand{\LeakagePower}{1.3\,mW}

\title{An 826\,MOPS, \avgPoweru{} Unum ALU in 65\,nm}

\author{
\IEEEauthorblockN{Florian Glaser, Stefan Mach, Abbas Rahimi, Frank K. G\"urkaynak, Qiuting Huang, Luca Benini}
\IEEEauthorblockA{ETH Zurich, Integrated Systems Lab IIS, Zurich, Switzerland \\
\{glaser, mach, rahimi, kgf, huang, benini\}@iis.ee.ethz.ch\vspace{-4mm}}
}

\begin{document}

\maketitle

\begin{abstract}
To overcome the limitations of conventional floating-point number formats, an interval arithmetic and variable-width storage format called universal number (unum) has been recently introduced~\cite{UnumBook}.
This paper presents the first (to the best of our knowledge) silicon implementation measurements of an application-specific integrated circuit (ASIC) for unum floating-point arithmetic.
The designed chip includes a 128-bit wide unum arithmetic unit to execute additions and subtractions, while also supporting lossless (for intermediate results) and lossy (for external data movements) compression units to exploit the memory usage reduction potential of the unum format.
Our chip, fabricated in a 65\,nm CMOS process, achieves a maximum clock frequency of \maxClkFreq{} at \TestVolt{} with an average measured power of \avgPoweru{}.
\end{abstract}

\section{Introduction}
\label{sec:intro}
Large scale data analytics and numerical applications have very widely ranging requirements in terms of numerical precision. 
While approximate computing shows flexibility with low precision arithmetic and aggressive bit width reduction~\cite{FP_AppComputing}, the other side of the application spectrum adheres to the IEEE standard for floating-point arithmetic~\cite{FP_HandBook} (IEEE~754) in spite of its possible side effects e.g., accumulation of rounding errors~\cite{FP_iisues} that can cause deviation from the exact value.  
To cover this wide range of demands, efficient hardware solutions that retain as much flexibility as possible, are highly desirable. 

The IEEE~754 format mainly suffers from rigid allocation of bits to its sign, exponent and mantissa fields and lacks robustness to rounding errors~\cite{Gustafson16}.
The latter weakness is caused by the implicit rounding rules defined in the standard:
When a desired value lies in between of two representable values, it will be forced to be rounded to the next best value producing an inevitable rounding error; across multiple calculations, such rounding error can be accumulated without allowing the application an explicit observation or control over the error.   
As an alternative, the universal number (unum)~\cite{Interview_Gustafson} format is proposed by John L. Gustafson to better control precision loss.
The goal of unum is to overcome the limitations of the IEEE~754 format by introducing a variable-width storage format, and a \emph{ubit} which determines whether a unum corresponds to an exact number or an interval between exact unums, hence explicitly representing when a calculation produces a value that is not exactly representable in the number system.
Therefore, the \emph{ubit} explicitly enables \emph{encoding} the error bound.
The unum format additionally defines two fields that make the number self-descriptive, as discussed briefly in Section~\ref{sec:back} and deailed in~\cite{UnumBook}.

The unum format, so far, has been supported in various programming environments including Julia~\cite{Julia}, Matlab~\cite{mnum}, Python~\cite{pynum}, J and Mathematica~\cite{Interview_Gustafson} languages.
Very recently, initial efforts on hardware with unum support focus on early synthesis~\cite{Unum_CEA_LETI} of three operators (i.e., addition, multiplication, and comparison), and FPGA implementation~\cite{Unum_FPGA} of four operators (i.e., addition, subtraction, multiplication and division).
To clearly evaluate the benefits and challenges of unum hardware design in silicon, we present -- to the best of our knowledge -- the first ASIC as a fully operational unum processor capable of performing additions and subtractions as well as format-specific functions for lossless and lossy compressions. This paper makes the following contributions:

\begin{itemize}
\item We present an ASIC integrating a unum arithmetic unit (ALU), supporting addition, subtraction,  implicit lossless and explicit lossy compression, measuring {\AluArea} in 65\,nm CMOS.
\item We report measurement results of the fabricated chip, achieving a maximum clock frequency of \maxClkFreq{} at \TestVolt{} with an average power of \avgPower.
\item We critically analyze advantages and shortcomings in supporting the unum format in hardware.
\end{itemize}

The rest of the paper is organized as follows. Section~\ref{sec:back} provides background on the unum format, how to perform computations with it and discusses associated advantages and shortcomings in terms of precision and memory footprint. Section~\ref{sec:asic} presents synthesis experiments for the IEEE~754 and unum compatible arithmetic units, followed by the design and optimization of the implemented ALU.
In Section~\ref{sec:exp}, we present the chip implementation and experimental results. Finally, Section~\ref{sec:conc} concludes the paper with a discussion of results.

\section{Unum Computing Background}
\label{sec:back}

\begin{figure}[b]
    \centering{\vspace{-7mm}
    \includegraphics[width=\linewidth]{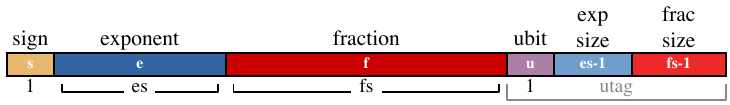}
    \caption{The unum format, extending \emph{sign-exponent-mantissa} floats with self-descriptive fields in the \emph{utag}.}
    \label{fig:unumfmt}}
\end{figure}

\subsection{The Unum Format}

\begin{figure}[t]
    \centering{
    \includegraphics[width=\linewidth]{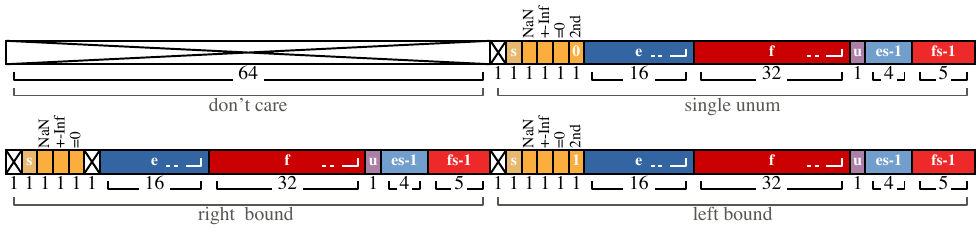}
    \caption{Layout of the internal representation of single unums (top) and ubound values (bottom) in the 128-bit register file.}
    \label{fig:unumfmt_hw}\vspace{-6mm}}
\end{figure}

The unum format, depicted in \figref{fig:unumfmt} bears similarity to the IEEE~754 floating-point representation for real numbers with its \emph{sign-exponent-mantissa} notation.
The unum format extends this representation by adding three new fields that allow for the inclusion of self-descriptive information about the represented value.
These additional fields are summarized under the name \emph{utag}. 

The last two fields in the utag denote the exponent size $es$ and fraction size $fs$ of the unum, making unum a variable-size format.

Hence, floating-point values that can be represented with a small number of bits require fewer storage bits compared to a large fixed-size floating-point environment thanks to the self-descriptive nature of the utag.

Since it is practically not feasible to allow for unlimited exponent and mantissa sizes, the widths of the exponent size and fraction size fields in the utag are fixed, defining the maximum range of possible unum values.

The chosen widths for the exponent size and fraction size fields then define a so-called unum \emph{environment}. For example, setting the exponent size width to 4~bits and the fraction size width to 5~bits, the resulting environment can represent unums with \emph{up to} 16 exponent and \emph{up to} 32 fraction bits. Such unums are defined in a \{4,5\}-environment -- the maximum possible size of a unum in an \{a,b\}-environment is given as $\mathrm{maxubits} = 2 + 2^a + 2^b + a + b.$

The first field in the utag, called the \emph{ubit}, can be set to denote that the represented value $x$ is not an exact point on the real line, but rather an open interval $\left(x,x+ulp\right)$ with $ulp$ being the unit in the last place for the current unum format. Explicitly encoding that the exact value cannot be represented in the current format sets unum apart from regular floating-point representations where all encoded values are considered as exact and approximation is completely implicit.

For describing general intervals more than one \emph{ulp} apart, two unums can be connected to create a so-called \emph{ubound}\footnote{This definition deviates from Gustafson's definition in \cite{UnumBook}, where the term \emph{ubound} can also denote a single unum with the \emph{ubit} set.}, each denoting one endpoint of an interval.
In a ubound, each of the two ubits indicates whether the respective endpoint is part of the interval or not, i.e., whether the interval is closed or open there.

\subsection{Unum Operations}

In this work's implementation, we include the basic operations that are addition and subtraction. 
Unum addition is similar to floating-point addition, with more complex special cases involving infinities being dependent on both values and bound types.
The left and right bound of ubounds can be handled independently, however.

One complexity of the floating-point arithmetic, namely rounding, is greatly simplified in unum: whenever the result of an operation on two exact values requires more precision than available in the unum environment, the ubit is set to mark the value as inexact. When handling bounds, the bound type of the result bound corresponds to the logical-OR of its operand ubits.

Since the bit-pattern representation of a value is not unique within a unum environment, there are additional unum-specific operations to be considered.
Since implementations should strive to utilize as little bits as possible for a given value, we also define the lossless \emph{optimize} operation, calculating the representation of a ubound with the smallest number of bits. 
Furthermore, Gustafson~\cite{UnumBook} specifies the \emph{unify} operation that attempts to merge a ubound consisting of two unums into the smallest single unum that fully includes the interval. This operation can incur loss of precision, namely if the resulting inexact unum covers a larger interval than the initial ubound.

\subsection{Considerations for Unum in Hardware}

The interchange format for unums as shown in \figref{fig:unumfmt} is specified in ~\cite{UnumBook}. Unum values reside in memory in this format, using only as much storage as mandated by the exponent size and fraction size fields -- which can be drastically less than using a fixed-width floating-point representation. This departure from using uniformly sized and aligned operands however requires additional effort when handling unums in the memory system. 

\begin{figure}[b]
    \centering{
    \vspace{-5mm}
    \includegraphics[width=\linewidth]{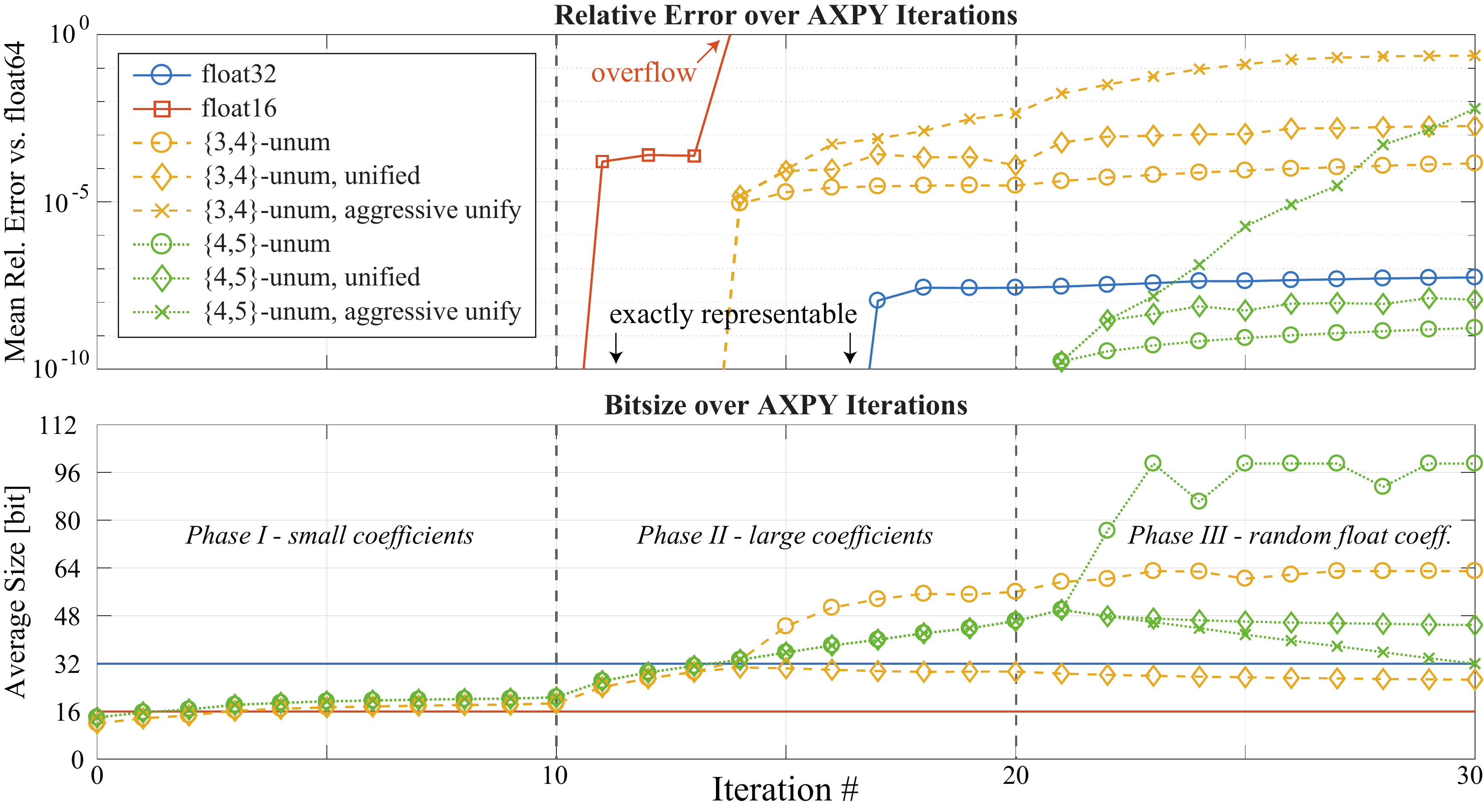}
    \caption{Relative error of \emph{axpy} iterations using floating-point and unum formats (top) and the bit-size of the results (bottom).}
    \label{fig:axpy}
    }
\end{figure}

In order to illustrate the dynamic behavior of unum during calculations, \emph{axpy} was run with input coefficients of rising complexity, calculating and accumulating the result using either floats or unum environments.
The change of the relative error compared to a double precision reference as well as the bit-size over the iterations is shown in~\figref{fig:axpy}.

\begin{figure}[t]
    \centering{
    \includegraphics[width=\linewidth]{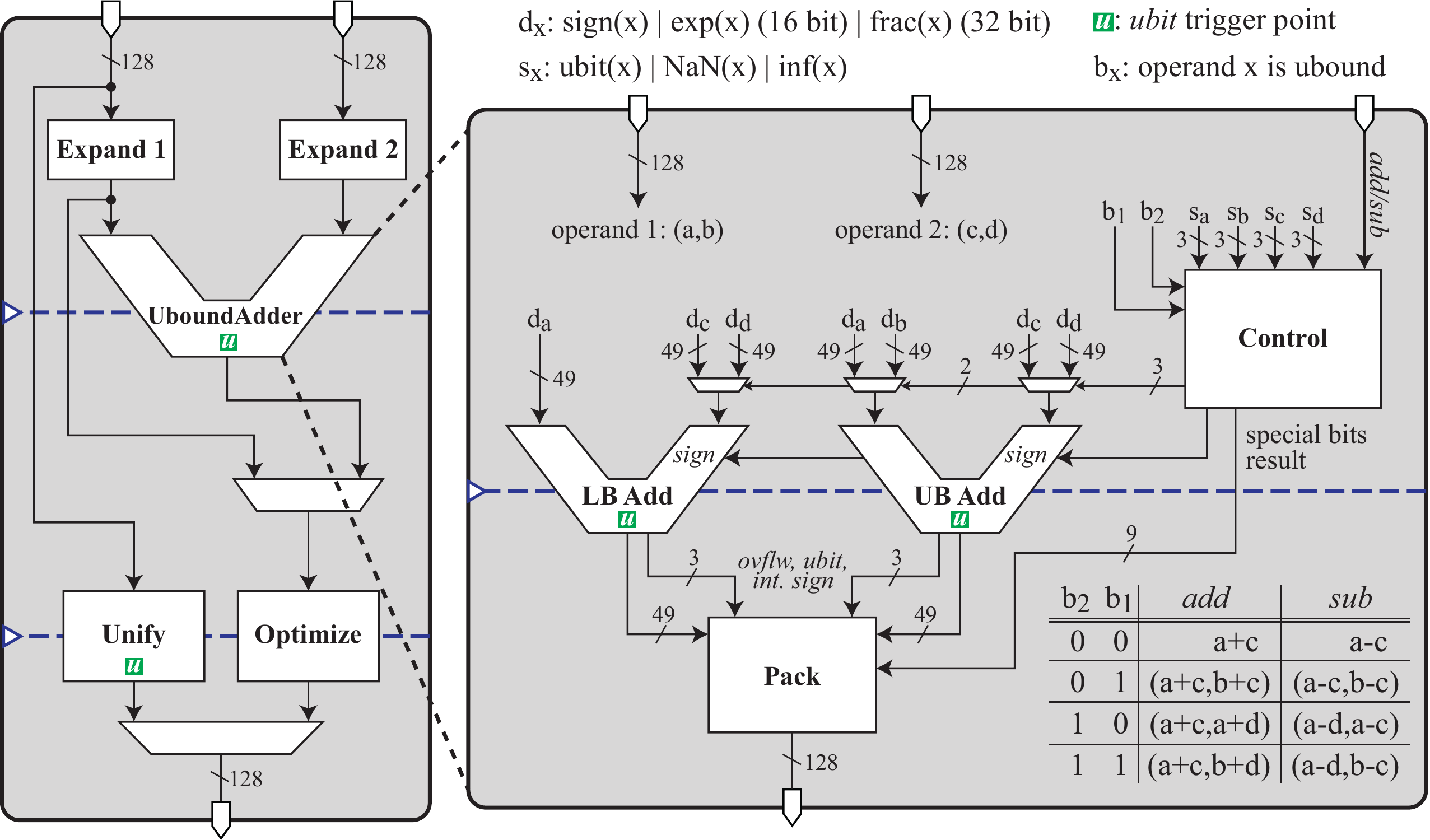}
    \caption{Data path of the proposed, 128-bit wide ALU and architecture of the unum adder along with supported operations. Blue lines indicate automatically retimed pipeline stages.}\label{fig:alu}\vspace{-5mm}}
\end{figure}

During phase I, only small coefficients are used, leading to results that can be exactly represented in all evaluated formats. The size of unum results is made up of the fixed size of the utag -- 8\,bit and 10\,bit, respectively, for the \{3,4\} and \{4,5\} environments -- and the dynamic number of bits needed to store the actual value.

Phase II applies large coefficients, significantly increasing the accumulated values. Unum formats start increasing in size to still accurately store the result. Once the exact value requires more fraction bits than available in the format, error proportional to the format-specific minimal \emph{ulp}-width appears and unum starts using ubounds to accurately represent the uncertainty of the results.

In phase III more error is introduced by using random floats as coefficients, causing also the \{4,5\}-unum's 32 fraction bits to be insufficient for exact results.

The ubounds used for unum results would require significantly more storage space than floats, thus they should stay contained within the processing unit registers if possible. Before storing to main memory, \emph{unify} can be used to reduce storage size at the cost of increasing the error bound. \emph{Unify}ing excessively, for example after each iteration as shown in \figref{fig:axpy}, causes the additional error introduced by each unification to quickly accumulate.

We notice in this example that there is a range where unum provides lower memory footprints than float32 with equivalent accuracy, while float16 error already grows rampant. \emph{Unified} \{3,4\}-unums require 7\% less memory than float32 at the price of a significant error increase similar to float16 -- while remaining usable long after float16 overflows due to insufficient range. \emph{Unified} \{4,5\}-unums require around 45\% more storage than float32 values mostly due to utag overhead -- albeit at around 5$\times$ lower error and explicitly denoting this error. Using float32 interval arithmetic to store the error bound would cost 39\% more memory compared to unum in this example.

Since arithmetic units and register files must be provisioned for handling all possible unums in a given environment, this incurs a relative hardware overhead for those unums that do not use the maximum width of the environment.

Unpacking of unum values in the register file and the storage of additional meta-information, called \emph{summary bits} in~\cite{UnumBook}, can simplify the implementation of unum operations, especially the handling of bounds and special cases such as NaN and infinity operands. As our ALU is targeted to extend embedded processing systems, we follow this approach in our implementation.

\section{Unum ALU Design}
\label{sec:asic}

\begin{figure}[b]
    \centering{\vspace{-4mm}
    \includegraphics[width=\linewidth]{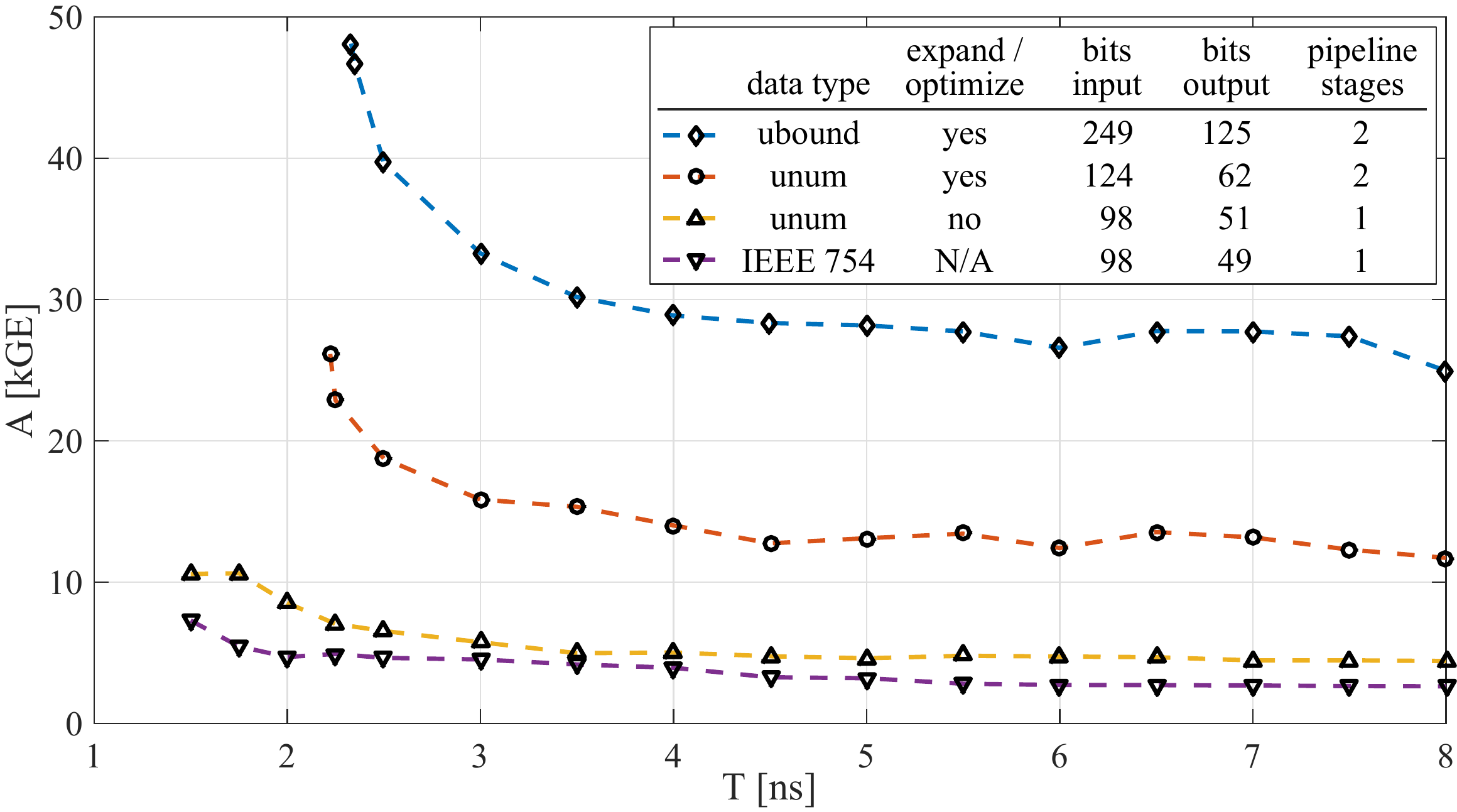}
    \caption{Area and timing comparison of the proposed ubound adder and its sub-parts against an IEEE~754 compliant adder.}\label{fig:AT}}
\end{figure}

We present a fully unum-\{4,5\} compatible ALU with support for a subset of the arithmetic and unum-specific operations proposed in \cite{UnumBook}. The design is targeted for integration into embedded parallel processing systems as a tightly memory-coupled accelerator, or a core data path extension. We thus follow the hardware-oriented unpacked data format for representing unums proposed in~\cite{UnumBook} to a large extent; details of the employed format are shown in \figref{fig:unumfmt_hw}. One single unum operand in this internal format is 64\,bit wide.

The maximum number of bits needed to represent these unums is $\mathrm{maxubits}=59$\,bits. We add the summary bits for \emph{NaN, $\pm\infty$, =0} as well as the \emph{2nd} flag to mark a unum part of a ubound, the ALU datapath that supports parallel operations for ubounds is therefore 128-bit wide.
\vspace{-1mm}

\subsection{ALU Architecture}
\renewcommand{\arraystretch}{1.2}

\begin{table}[t]
\begin{center}
\caption{Post-layout area distribution of the proposed ALU}\label{tbl:area_dist}
\begin{tabular}{@{}lr@{}}
\toprule
\textbf{Overall ALU area} & \textbf{50\,kGE / 0.07\,mm$^2$} \\ \hline
Lower, upper bound adder, each & 14\,\% \\
Expand units, each & 17\,\% \\
Unify unit & 27\,\% \\
Optimize unit & 7\,\% \\
Control, data routing & 6\,\% \\
\bottomrule
\end{tabular}
\vspace{-7mm}
\end{center}
\end{table}
The ALU is depicted in \figref{fig:alu} and can perform additions and subtractions on either two ubounds, two unums or one ubound and one unum.
Additionally, the format-specific functions \textit{optimize} and \textit{unify} were implemented: With \textit{optimize}, lossless compression is provided on the one hand by calculating the representation with the smallest exponent and fraction size for a given unum or ubound. On the other hand, the (usually) lossy \textit{unify} reduces a ubound to a unum whenever possible, saving potentially half of the storage at the expense of precision.
Consequently, the adder and the \textit{unify} unit can possibly output inexact results from exact operands; this behavior is deeply manifested in the unum format by a set \textit{ubit}. All units with the capability of introducing this format-specific number property are marked with an inverted, green~\textit{u} in \figref{fig:alu}.
\vspace{-1mm}

\subsection{Unum Adder}

\begin{figure}[t]
    \centering{
    \includegraphics[width=0.7\linewidth]{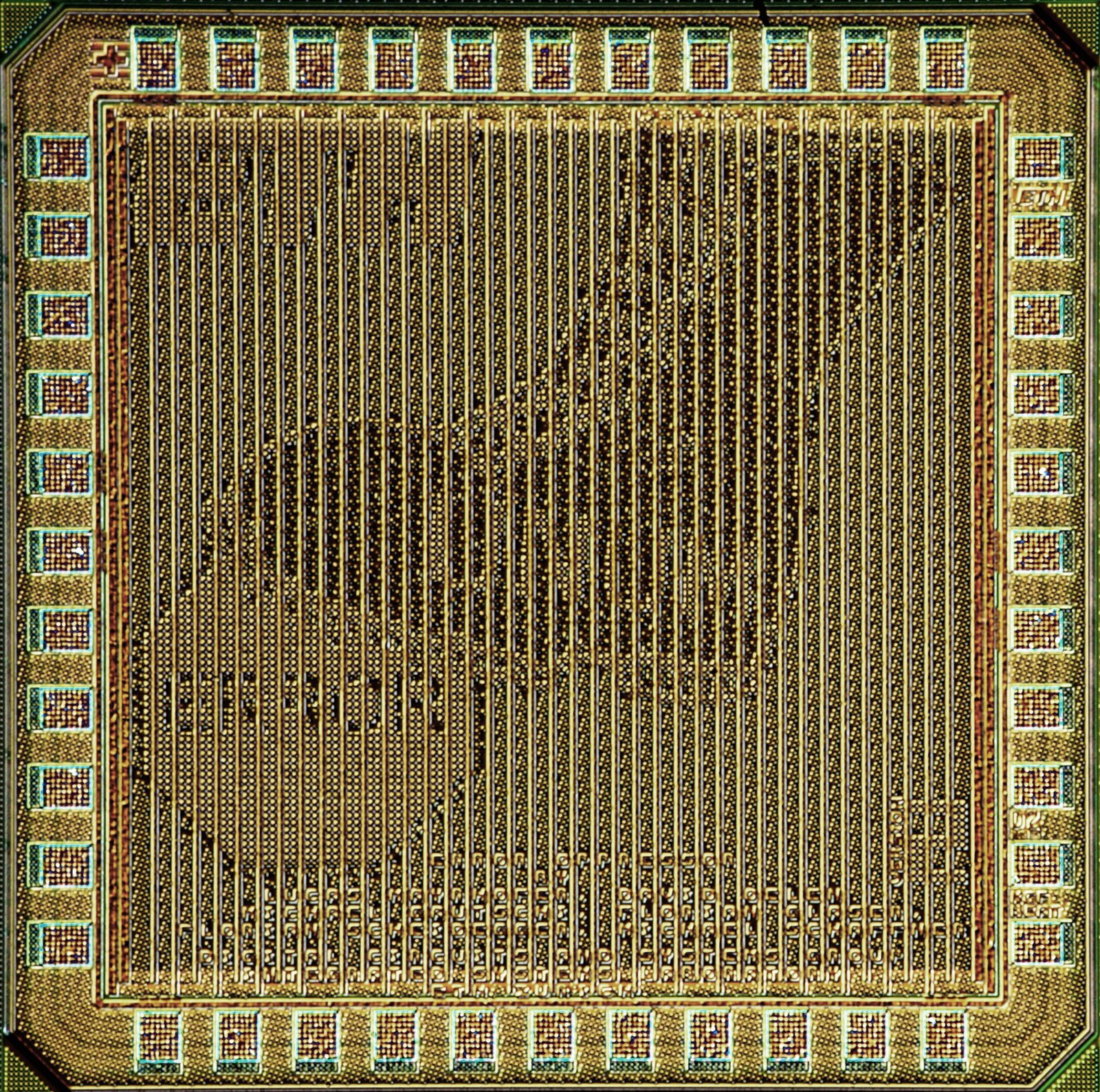}
    \caption{Die micrograph of the taped-out ASIC.}\label{fig:chip}
    \vspace{-5mm}
    }
\end{figure}

The adder is internally split into separate data paths for the calculation of the resulting upper and lower bounds in case any of the operands is a ubound. The operands are denoted as $(a,b)$ and $(c,d)$ in case of ubounds and $a$ and $c$ in case of unums, respectively. In order to take advantage of the regularity of the floating-point arithmetic units, the operands are expanded to the maximum supported precision with 16 exponent and 32 fraction bits beforehand. The core of each adder then consists of a floating point adder of appropriate size with hidden bit, overflow and rounding support, complemented with checks for unum infinity, zero and NaN special cases. Most importantly, however, the adder detects if its result cannot be represented exactly and sets the \textit{ubit} in such cases.

\subsection{Optimized Compression}

Particular focus was put on optimizing the routing of data through the available compression units during ALU design: The \textit{optimize} operation is carried out both through a dedicated opcode as well as implicitly after every adder operation to leverage the storage-saving capabilities of the unum format; the \textit{unify} operation can only be carried out with an explicit opcode to maintain controllability over all lossy operations. In a typical processing environment, intermediate results can then be successively \textit{optimized} while \textit{unified} only once and right before expensive data movements, e.g., DRAM transfers. This mechanism allows for maximum storage savings while not sacrificing desired intermediate precision.

\subsection{Comparison with IEEE~754}

\figref{fig:AT} shows synthesis experiments in 65\,nm, comparing different unum-enabled arithmetic units with an IEEE~754 compliant floating-point adder with corresponding exponent and fraction sizes. A first observation is a modest area increase (27\% or 1.08\,kGE with a 4\,ns period constraint) when only considering the unum adder. However, complementing the adder with the expand and \emph{optimize} units to take advantage of on-the-fly data compression comes with an area increase of more than 3.5$\times$. The implemented, fully-parallel ubound adder adds roughly another factor of two while also providing double the throughput. The second important observation is the limitation in terms of minimum clock period for the compression-enabled unum units, even with an additional pipeline stage.
Table~\ref{tbl:area_dist} confirms the findings that compression-related blocks consume a significant part of the overall ALU area; they however can be reused and shared between arithmetic operations.

\section{ASIC Implementation}
\label{sec:exp}

For silicon verification and characterization, we embedded the proposed ALU into a test-bed consisting of instruction SRAM, register file and control state machine. A maximum of 1024 instructions can be executed sequentially once or repeatedly, hiding IO delays to emulate operating conditions resulting from integration into embedded processing systems.

\subsection{Experimental Setup}

Both SRAM and register file are accessible for writing and reading through dedicated commands to a memory controller block; consequently, the maximum ALU speed can be determined after preloading instruction memory and register file with suitable instructions and data, respectively. Results from the register file are then read out and verified against a golden model implementation \cite{pynum}. The design nets \ChipArea{} of circuit area within the ASIC die pictured in \figref{fig:chip}.

\begin{table}[b]
\begin{center}
\caption{Measured characteristics of unum-\{4,5\} ASIC, all numbers acquired from measurements at \TestVolt{} at room temperature}\label{tbl:chip_results}
\begin{tabular}{@{}llr@{}}
\toprule

Technology / Supply && umcL65 / 1.2\,V \\ 
Circuit Area && \ChipArea \\ \hline
Measured Leakage Power && \LeakagePower \\ 
Measured Dynamic Power && \avgPower \\ \hline
\multirow{3}{*}{Maximum Speed} & Add/Subtract      & 413\,MHz \\
               & Unify    & 468\,MHz \\
               & Optimize & 471\,MHz \\
 
\bottomrule
\end{tabular}
\end{center}
\end{table}

\subsection{Experimental Results}

The fabricated prototypes were characterized on a commercial \emph{Advantest SoC V93000} ASIC tester, using full-range data  generated in a directed random fashion. The findings with further ASIC properties are summarized in Table \ref{tbl:chip_results}.

\section{Conclusion}
\label{sec:conc}

We presented measurement results of the first unum-\{4,5\} ALU ASIC implementation.
Our 128-bit wide ALU supports addition and subtraction of ubounds and the unum-specific operations \textit{optimize} and \textit{unify} at up to \maxClkFreq{}, allowing up to 826\,M unum additions or subtractions per second. We discussed synthesis experiments for the comparison of unum-enabled arithmetics with the IEEE~754 counterparts and conclude that it must be carefully analyzed whether memory accesses are expensive enough for the significant (de)compression overhead linked to variable-width number formats to pay off. 
Furthermore, we touched on the possible storage-saving capabilities of the unum format through an example, concluding that unum formats provide moderate memory footprint advantage (7\%) with respect to the standard FP32 and wider range than FP16, at a price of a significant increase in datapath complexity and requiring special care in avoiding aggressive unification to prevent error blow-up. 

\section*{Acknowledgments}

The authors gratefully acknowledge the support of David Oelen and Lucas Mayrhofer during ASIC design and testing.

\bibliographystyle{IEEEtran}

\bibliography{IEEEabrv,bibliography}

\end{document}